\documentclass[twocolumn,preprintnumbers,amsmath,amssymb]{revtex4}

\usepackage{graphicx} 
\usepackage{dcolumn}  
\usepackage{bm}       
\usepackage{color}    

\begin{document}

\title{Ellipticity Dependence of Excitation and Ionization of Argon Atoms by Short-Pulse Infrared Radiation}
\author{Thomas~Pauly$^1$, Aaron Bondy$^{1,2}$, Kathryn R. Hamilton$^1$, Nicolas~Douguet$^{3}$, 
        Xiao-Min Tong$^4$, Dashavir Chetty$^5$, and K.~Bartschat$^1$}
\affiliation{$^1$Department of Physics and Astronomy, Drake University, Des Moines, Iowa 50311, USA}
\affiliation{$^2$Department of Physics, University of Windsor, Windsor, Ontario N9B 3P4, Canada}
\affiliation{$^3$Department of Physics, Kennesaw State University, Kennesaw, Georgia 30144, USA}
\affiliation{$^4$Center for Computational Sciences, University of Tsukuba, 1-1-1 Tennodai, Tsukuba, Ibaraki 305-8573, Japan}
\affiliation{$^5$Centre for Quantum Dynamics, Griffith University, Brisbane, Queensland 4111, Australia}

\date{\today}
\begin{abstract}
When atoms or molecules are exposed to strong short-pulse infrared radiation, 
ionization as well as ``frustrated tunneling ionization'' (FTI) 
can occur, in which some of the nearly freed electrons recombine into the initial ground
or an excited bound state. We analyze the ellipticity dependence of the relative
signals that are predicted in a single-active electron approximation (SAE), the validity of which is
checked against a parameter-free multi-electron \hbox{$R$-matrix} (close-coupling) with 
time dependence approach.
We find good agreement between the results from both models, thereby providing confidence in the SAE
model potential to treat the process of interest.  Comparison of the relative excitation probabilities
found in our numerical calculations with the predictions of Landsman {\it et al.}\ 
(New Journal of Physics {\bf 15} (2013) 013001) and Zhao {\it et al.}\ (Optics Express {\bf 27} (2019) 21689)
reveals good agreement with the former for short pulses.  For longer pulses, the ellipticity
dependence becomes wider than that obtained from the Landsman {\it et al.} formula, but we
do not obtain the increase compared to linearly polarized radiation predicted by Zhao {\it et al.}
\end{abstract}

\maketitle

\section{Introduction}

Strong-field ionization utilizes a slowly varying few-cycle infrared laser 
pulse to eject an electron from an atom. With an intense infrared laser, the 
electric field significantly alters the effective potential. In this process described 
by Corkum~\cite{Corkum3Step}, the electron has the ability to ``tunnel'' through the effective 
barrier and become free.  After ionization, however, the electric field can guide the 
electron back to the target ion. There it can be recaptured into the ground 
state or any other bound state, such as high-lying \hbox{Rydberg} states. This 
is the strong-field process of ``frustrated tunnel ionization'' (FTI), which is negligible in 
the treatment of the standard photoeffect. 
From some of these excited bound states, 
the electron can jump to metastable states of the atom.  Production of metastable 
sources through FTI may have practical advantages, for example, a 
potentially high yield with minimal heating of the sample~\cite{Dakka2018}.

In addition to studying the process with linearly polarized light, the ellipticity dependence
of both FTI and successful ionization has been of significant theoretical interest.  Intuitively, 
one might expect that FTI will diminish with increasing ellipticity of the
IR radiation, since rescattering is most likely a very important contributing process, in addition
to direct excitation.  Successful ionization, on the other hand, while also potentially benefitting
from recollisions, may be less sensitive to the latter. If these hypotheses are correct,
one would expect the rates for these processes, at a fixed peak intensity of a pulse (in practice,
a fixed total energy delivered by the pulse), to peak at or near zero ellipticities, i.e., linearly
or (near-)linearly polarized light.  It should then drop off with increasing ellipticity and
have a minimum for (near-)circularly polarized light.  Also, the drop should be steeper for FTI than for
actual ionization, as discussed in~\cite{PhysRevA.100.063424}.

An early FTI experiment was performed by Nubbemeyer {\it et al.}~\cite{PhysRevLett.101.233001}
on helium, where the above expectations were, indeed, fulfilled.  Since FTI is a strong-field
process, in which the motion of the electron is predominantly driven by the slowly varying electric
field,  it is not surprising that models were developed in which the motion of the electron
was described classically, with very limited account for the target structure.
Landsman {\it et al.}~\cite{LanNJP13}, for example, derived an analytical formula 
based on the strong-field approximation model, which neglected the effect of the Coulomb field until the pulse has passed. 
They considered the conditions where the initial transverse momentum of the tunneled electron is cancelled 
by the drift momentum gained in the field. This results in a \hbox{Gaussian} probability distribution with 
respect to ellipticity, which depends only on the laser intensity, wavelength, and ionization potential 
of the target. The approximations employed in the derivation of the formula limit it, in principle, to laser fields 
much stronger than the Coulomb field and small $\varepsilon$, but  it was apparently 
sufficient to reproduce the experimental results of Nubbemeyer {\it et al.}~\cite{PhysRevLett.101.233001}.
More recent work on the helium target was reported by Yun {\it et al.}~\cite{Yun2018}.

For situations involving targets with lower ionization potentials or lower laser intensities, semi-classical 
calculations that account for the Coulomb field even during the laser interaction, such as those  
presented by Shvetsov-Shilovski {\it et al.}~\cite{Shvetsov-Shilovski2009}, are typically employed. 
At lower intensities, these calculations sometimes predict a maximum in the excitation probability at a non\-zero ellipticity. 
Zhao {\it et al.}~\cite{Zhao2019}, for example, recently reported predictions for FTI of Ar($3p$) with laser pulses of
two wavelengths (800 nm and 1,600 nm), two peak intensities ($\rm 0.8 \times 10^{14}\,W/cm^2$  and
$\rm 1.6 \times 10^{14}\,W/cm^2$, and pulse\-lengths of 30, 10, and~4 cycles, respectively.
At the higher intensity, the distribution deviates from a Gaussian shape, 
exhibiting a relatively flat top around linear polarization $\varepsilon=0.0$.
At the lower intensity and a pulse length of 30~fs for the 800~nm radiation, a maximum of the 
relative FTI rate occurs at $\varepsilon=0.2$,
where there is a predicted increase of more than 10\% compared to  $\varepsilon= 0$.
For shorter pulses of 10 and 4 cycles, the increase disappears in the Zhao {\it et al.}~\cite{Zhao2019} predictions.  
We note that peaks at $\varepsilon\ne 0$ were previously predicted 
for Mg~\cite{PhysRevA.87.033415}, where it was 
shown that they are sensitive to the initial conditions of the calculations. 

No experimental data to confirm or disprove the Zhao {\it et al.}~\cite{Zhao2019} predictions are currently
available, but an experiment is underway at Griffith University.  Preliminary results were
already announced~\cite{ICPEAC-Griffith}, but no firm conclusions are available yet.  In fact, further 
theoretical guidance is highly desirable for this challenging experimental project.

Very recently, experimental FTI and ionization results for Ar($3p$) obtained with linear polarization 
were published~\cite{fti_linear_ar}.
These data were compared with calculations carried out by the authors of the present paper. Good qualitative
agreement was obtained with predictions from a single-active-electron (SAE) model, with the effective potential 
proposed by Tong and Lin~\cite{Tong05c}.  A few calculations to cross-check those results were also
carried out with other potentials proposed in the literature, and -- most importantly -- with predictions from a
fully {\it ab initio} all-electron \hbox{$R$-matrix} with time dependence (RMT) approach. The latter (see below) is
based on the close-coupling expansion, with full account for electron exchange where it is potentially important  and
a multi-configuration Hartree-Fock target description.

The principal objectives of the present work were to extend those latter calculations to elliptically
polarized light.  This would allow us to i) further test the validity of the SAE approach by comparing with RMT predictions,  
ii)~once again check the predictions of Landsman {\it et al.}~\cite{LanNJP13}
and Zhao {\it et al.}~\cite{Zhao2019} (which clearly disagree for the 800~nm, 30~fs, $\rm 0.8 \times 10^{14}\,W/cm^2$ case), 
and iii)~provide guidance for the experimental investigations.  In the latter context, it is also of interest to determine
how the experimental signal might actually arise. The measured observable is the relative yield of metastable  
Ar atoms in the $(3p^5 4s)^3P_2$ and $(3p^5 4s)^3P_0$ states.  Given that the ground state is a relatively well
$LS$-coupled singlet state, while the meta\-stables are equally well $LS$-coupled triplet states, the 
process almost certainly proceeds via intermediately-coupled excited states, whose triplet component 
can be reached while fulfilling the spin-conserving selection rule of electric dipole radiation.

To model the actual experiment, therefore, one would ideally carry out a semi-relativistic
calculation accounting for the intermediate-coupling nature of the target states involved, followed
by a set of rate equations to predict the ultimate fate of the $3p$ electron that is subjected to the
strong laser field.  Even though the RMT method has recently been extended to the semi-relativistic
regime~\cite{spin-orbit-rmt}, such calculations for the process of interest here are far beyond the currently 
available computational resources.  Fortunately, as shown in~\cite{fti_linear_ar},  
the {\it relative} probabilities as a function of laser parameters appear to be predictable
without these efforts, and hence the present calculations are expected to be useful in the further planning 
of the experimental study.

This paper is organized as follows. In section~\ref{sec:comp} we briefly describe the 
SAE and RMT models used in the present calculations.  This is followed by our results presented 
in section~\ref{sec:results}, with the conclusions drawn in section~\ref{sec:conclusions}.
Unless specified otherwise, atomic units are used throughout.

\section{Computational Details}\label{sec:comp}

\subsection{Single-Active-Electron Approach}
The basic principles behind these calculations were described in~\cite{fti_linear_ar}.  
In modeling the process, we need to solve the non\-relativistic time-dependent Schr\"odinger equation (TDSE).
In atomic units, this is a partial differential equation
\begin{equation}
{\rm i}\frac{\partial}{\partial t} \Psi(\bm{r},t) = [H_0 + V(t)]\,\Psi(\bm{r},t)
\label{eq:TDSE}
\end{equation}
for the wave~function~$\Psi(\bm{r},t)$.  Here $H_0$ is the field-free Hamiltonian and $V(t)$ is the
time-dependent interaction represented by the laser pulse.  Instead of including the kinetic
energy and the Coulomb interaction of all electrons with the nucleus and with each other, only one electron is 
considered in the
Single-Active-Electron (SAE) approximation. Its interaction 
with the nucleus and all other electrons is approximated by a (usually local) potential.
For the present work, we used the potential proposed by Tong and Lin~\cite{Tong05c}), which was recently employed
successfully in FTI calculations with linearly polarized radiation~\cite{fti_linear_ar}.
For maximum internal consistency, we recalculated the $3p$ and all other orbitals in this potential, and
we also used it to generate the distorted Coulomb waves needed to extract the ionization probability.

The electric field of the laser is represented by the vector potential
\begin{equation}
\bm{A}(t) = f(t)\,\frac{E_0}{\omega\,\sqrt{1 + \varepsilon^2}}\,\left[\cos(\omega t + \varphi)\,\hat{\bm{x}}
    + \varepsilon \sin(\omega t + \varphi)\,\hat{\bm{y}}\right].
    \label{eq:Pulse}
\end{equation}
Here $\varepsilon$ is the ellipticity, with $\varepsilon = 0$ corresponding to linearly polarized
and $\varepsilon = 1$ to circularly polarized light. Furthermore, $E_0$ is the maximum amplitude of the
electric field, and $\varphi$ is the carrier-envelope
phase (CEP). The factor $E_0/\sqrt{1 + \varepsilon^2}$ ensures that for a given pulse represented by
the envelope function~$f(t)$ (typically a Gaussian or a $\sin^2$ function closely resembling a Gaussian),
the same amount of energy (i.e., the field squared integrated over the length of the pulse) is
delivered to the target. This is needed to unambiguously test the effects of different ellipticities.
In order to avoid a potentially unphysical displacement~\cite{PhysRevA.90.043401},
we set the vector potential and calculate the electric field as its time derivative according to 
$\bm{E}(t) = -\frac{\displaystyle d}{\displaystyle dt} \bm{A}(t)$.

In writing Eq.~(\ref{eq:Pulse}), we chose the quantization axis ($\hat{\bm{z}}$) perpendicular to the electric field, 
i.e., along the laser propagation direction. For the special case of linear polarization ($\varepsilon = 0$),
it is better to rotate the coordinate system and take advantage of the cylindrical symmetry by choosing the
quantization axis parallel to the direction of the electric field vector.

Since the initial state is spherically 
symmetric, right-and left-hand elliptically polarized light yield the same answers.  In
the electric dipole approximation, there is no spatial dependence in 
$\bm{A}(t)$ and $\bm{E}(t)$. We use $\sin^2$ envelopes
with FWHM of the intensity set to 6~fs or 30~fs, respectively. 
For 800~nm, the FWHM time in~fs is very close to the number of cycles for a $\sin^2$ envelope of the field.
In the 6~fs calculations, we average over the CEP
by performing calculations for $\varphi = 0^\circ,~45^\circ,~90^\circ$ and $135^\circ$.

The initial state of the system (a $3p$ orbital with a magnetic quantum number~$m$) is propagated 
in time until the end of the pulse. In order to simulate an unpolarized $3p$ electron in the Ar atom, calculations are 
carried out for all possible initial projections of the orbital angular momentum, i.e., $m = 0,\pm 1$, and the results are then
summed incoherently.  
For linearly polarized radiation, the results for initial
$m = +1$ and $m= -1$ are the same and the value of $m$ is conserved.  As mentioned above, we then choose the quantization axis along 
the direction of the electric field.  In this case, we use the straight\-forward Crank-Nicolson (CN) method~\cite{CN1996}
with the electric dipole operator written in the length gauge to propagate the wave function.

The case of a non\-zero ellipticity, on the other hand,
is much more numerically challenging due to the reduced symmetry that
no longer conserves the initial projection of the angular momentum. Here we choose the
quantization axis perpendicular to the electric field, i.e., along the laser propagation direction, as written in Eq.~(\ref{eq:Pulse}).
Since the CN method is no longer applicable directly, we employ the matrix iterative method (MIM)~\cite{NF1999} with the
dipole operator chosen in the velocity gauge. 

After the pulse is over, the wave function of the active electron is projected onto the initial $3p$ orbital to obtain
the survival probability~$P_{\rm surv}$ as the square of the overlap matrix element, as well 
as onto selected excited states to yield individual excitation probabilities and finally on distorted
Coulomb waves to obtain the energy-differential ionization probability $dP/dE$, where $E$ is the energy of the continuum electron.
The integral over all ejected energies yields the total ionization probability~$P_{\rm ion}$.  Finally, the total probability 
for excitation~$P_{\rm exc}$ (regardless of the Rydberg state) is obtained by using the 
conservation of the norm of the total wave function, i.e.,
$P_{\rm exc} = 1 - P_{\rm surv} - P_{\rm ion}$.  Special care was taken to ensure that the numerics were sufficiently 
stable to ensure meaningful results, even when $P_{\rm surv}$ and $P_{\rm ion}$ are small. 

All the SAE calculations reported below were carried out with an updated version of the computer code described by
Douguet {\it et al.}~\cite{PhysRevA.93.033402} and references therein.  Also, spot checks against results from the 
independent SAE code used for most of the calculations reported in~\cite{fti_linear_ar} were performed.

\subsection{$\bm{R}$-matrix with Time Dependence Approach}

The $R$-matrix with time-dependence method (RMT) is an {\it ab initio}, multi-electron 
method capable of describing the interaction of general atoms and molecules with arbitrary 
laser fields~\cite{rmt_cpc,rmt_clarke,rmt_moore}. RMT employs the well-known $R$-matrix 
paradigm \cite{burke} of dividing configuration space into two separate regions, in this 
case over the radial coordinate of the ejected electron.  In the inner region (close to 
the nucleus) the time-dependent multi-electron wave function is represented by an 
$R$-matrix basis with time-dependent coefficients. This configuration-interaction 
approach ensures that we take into full account all electronic interactions, 
including electron correlation and exchange.  Specifically, we used the target description developed
by Burke and Taylor~\cite{rmt_argon_structure} for photo\-ionization of argon.

In the outer region, where one 
electron is sufficiently far removed that we can neglect the effect of exchange, 
the wave function is described in terms of residual-ion states coupled with the 
radial wave function of the ejected electron, and is expressed explicitly on a 
finite-difference grid. The outer region also includes several long-range potentials, 
which describe the interactions between the ejected electron, laser and residual ion. 

In contrast to traditional $R$-matrix-based approaches, the wave function itself is 
matched explicitly at the boundary, in both directions, rather than via an $R$ matrix. 
The wave function is propagated in the length gauge, as it has been found to converge 
faster than the velocity gauge with the atomic structure description typically employed 
in time-dependent $R$-matrix calculations~\cite{tdrm_dipole_gauge}.  Time propagation of 
the wave function is achieved by the Krylov subspace method of Arnoldi and Lanczos~\cite{arnoldi}. 
RMT has been successfully applied to a wide range of strong-field problems, from High Harmonic 
Generation (HHG) in two-color~\cite{hamilton_two_colour} and near-IR fields~\cite{ola_nearIR}, 
XUV-initiated HHG~\cite{brown_xuvhhg,clarke_xihhg}, and strong-field rescattering~\cite{ola_rescattering}.

The default polarization plane in the RMT code is the $zy$-plane, 
with $\hat{\bm{z}}$ and $-\hat{\bm{y}}$  the major and minor axes of the polarization ellipse, respectively. A general, 
elliptically polarized electric field will therefore be of the form:

\begin{equation}\label{def:efield}
   \bm{E}(t) = f(t)\,{\rm Re}\left[\hat{\bm{e}}
    \;{\rm e}^{-{\rm i}(\omega t + \varphi)}\right],
\end{equation}
where  $\hat{\bm{e}}=(\hat{\bm{z}} - {\rm i}\varepsilon \hat{\bm{y}})/\sqrt{1+\varepsilon^2}$ 
is an arbitrary polarization vector and Re[$x$] denotes the real part of the complex quantity~$x$.  
As before, $\varepsilon$ is the ellipticity and  
$f(t)$ is the envelope function, chosen to create a $\textnormal{sin}^2$ pulse envelope, 
and $\varphi$ is a carrier-envelope phase.  This choice of polarization axes allows 
calculations for linear fields ($\varepsilon = 0$) to be polarized along the $z$-axis. 
For calculations with circularly polarized laser pulses, it is often practical to 
rotate the polarization plane to the $xy$-plane, as this reduces the number of 
dipole-accessible states relative to the $zy$-plane, thereby reducing the size of the 
calculation significantly.

 In the current study, the argon target is described within an $R$-matrix inner region of 
 radius 20 and an outer region 
 of 3,500. The latter ensures that no significant part the wave packet describing the ejected electron 
 gets to the boundary, where it would be reflected. 
 The finite-difference grid spacing 
 in the outer region is 0.08 and the time step for the wave function 
 propagation is 0.01 ($=0.24$ atto\-seconds).  The description of argon includes all 
 available $3s^23p^5\,\epsilon\ell$ and $3s3p^6\,\epsilon\ell$ channels 
 up to a maximum total angular momentum of $L_{\mathrm{max}}=80$. 
 Doubly-excited states of the residual Ar$^+$ ion are also included to 
 ensure an accurate description of the \hbox{$3s3p^6np$} window 
 resonances~\cite{LEVY19691349}. 
 
 The continuum functions 
 are constructed from a set of 50~$B$-splines, of order 13, for each
available angular momentum of the outgoing electron.
  The ground-state survival probability is determined directly from 
  the wave function at the end of the laser pulse. The ionization 
  probability is calculated by firstly projecting the channel-resolved wave functions, 
  from some radius $r_{\rm min} \geq 20$  up to $r_{\rm max} = 3,500$, onto plane waves 
  to determine the photo\-electron spectrum, and then integrating over all energies. 
  As with the SAE method outlined above, the total excitation probability is then 
  obtained using the conservation of the norm of the wave function. 
 
Due to the significant computational expense of these calculations, 
RMT results are only available for select ellipticities at a laser 
intensity of $\rm 1.0 \times 10^{14}\,W/cm^2$.  In principle, the coupled-channel
formulation used here also allows ionization leaving the residual Ar ion in the
$(3s 3p^6)^2S$ state. However, this excited ionic state has a much higher ionization potential, and hence
the signal for the process of interest here is negligible.

\section{Results and Discussion}\label{sec:results}

\begin{figure}[t!]
\includegraphics[width=0.45\textwidth]{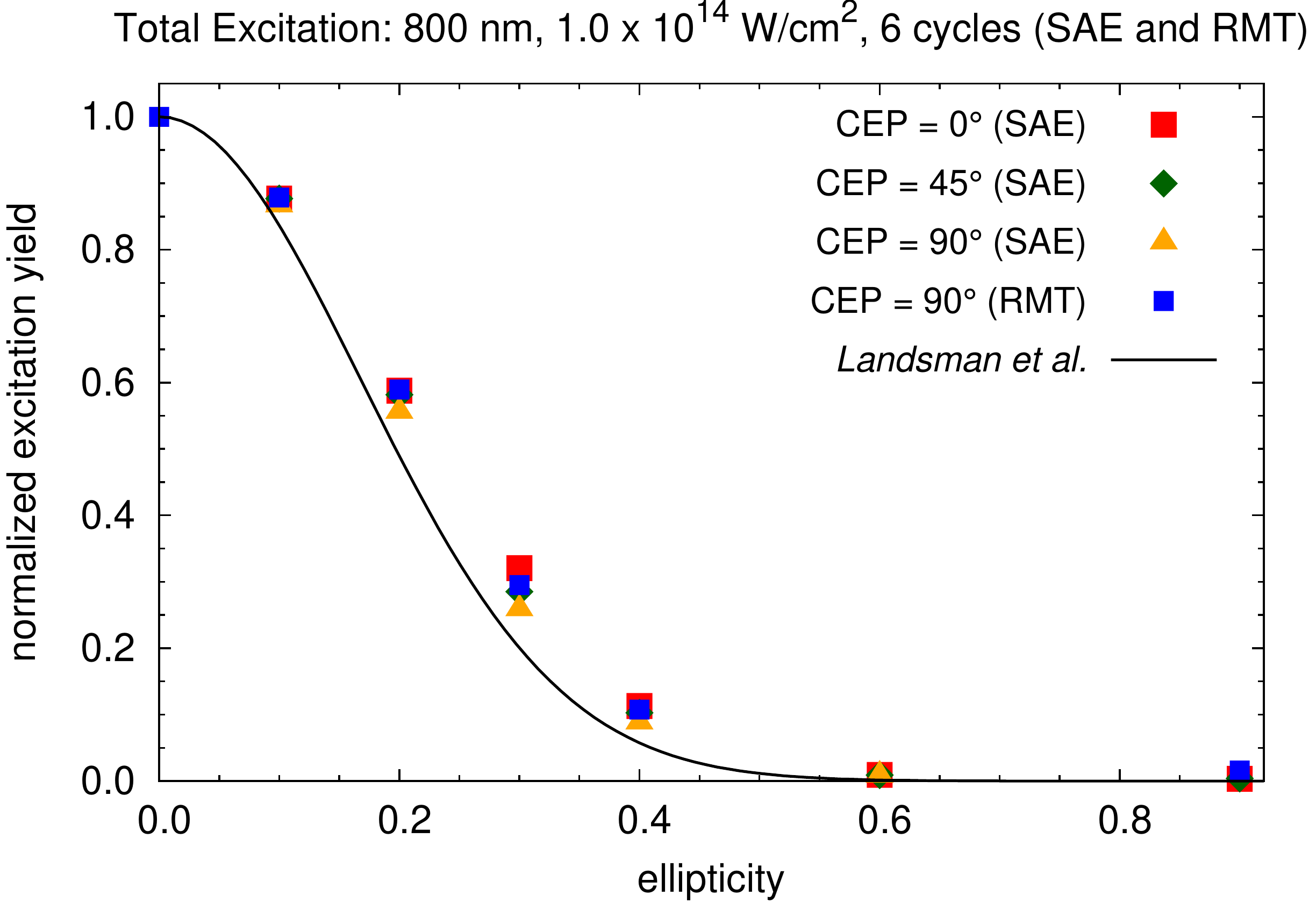}
\caption{\label{fig:fig2} Present SAE results for total excitation of Ar($3p)$ for a 6-cycle 800~nm
pulse of peak intensity $\rm 10^{14}\,W/cm^2$ for three different CEPs, compared with the 
predictions obtained from the Landsman {\it et al.} formula and an RMT calculation for a CEP of $90^\circ$.
See text for details.}
\end{figure}

Figure~\ref{fig:fig2} depicts a comparison of SAE results for three different CEPs of a 6-cycle 800~nm
pulse of peak intensity $\rm 10^{14}\,W/cm^2$.  Also shown are results from our RMT calculation and the predictions 
obtained from the Landsman {\it et al.} formula.  Recall that the electric field rather than the vector potential 
is set in the RMT code, and so the RMT calculation was carried out for an electric field with CEP = $0^\circ$, 
which is equivalent to an SAE calculation with CEP = $90^\circ$, assuming that the derivative of the envelope 
function only has a minor effect on
the calculation of the field used in the SAE calculation.  To simplify the comparison, we therefore use the label $90^\circ$
for the RMT calculations.  We also checked that SAE calculations performed with setting the electric field directly
(i.e., as in RMT), while producing an non\-zero displacement and small changes in the individual probabilities,
hardly affect the ratios and hence the conclusions of the present work.
Clearly, all the results are so close to each other that 
the all-electron RMT results for the observable of interest can certainly be taken as support for the much simpler
SAE calculations.

Generally, our results for the 6-cycle pulses are in qualitative agreement
with the  Landsman {\it et al.} formula, even though they assumed ``long'' pulses in their derivation and made 
further approximations mentioned in the Introduction. 
However, we note a small but systematic trend of the numerical results 
always lying slightly above those predicted by the formula. Finally, averaging over the CEP only has a small effect on the 
6-cycle results.  

\begin{figure}[h!]
\includegraphics[width=0.45\textwidth]{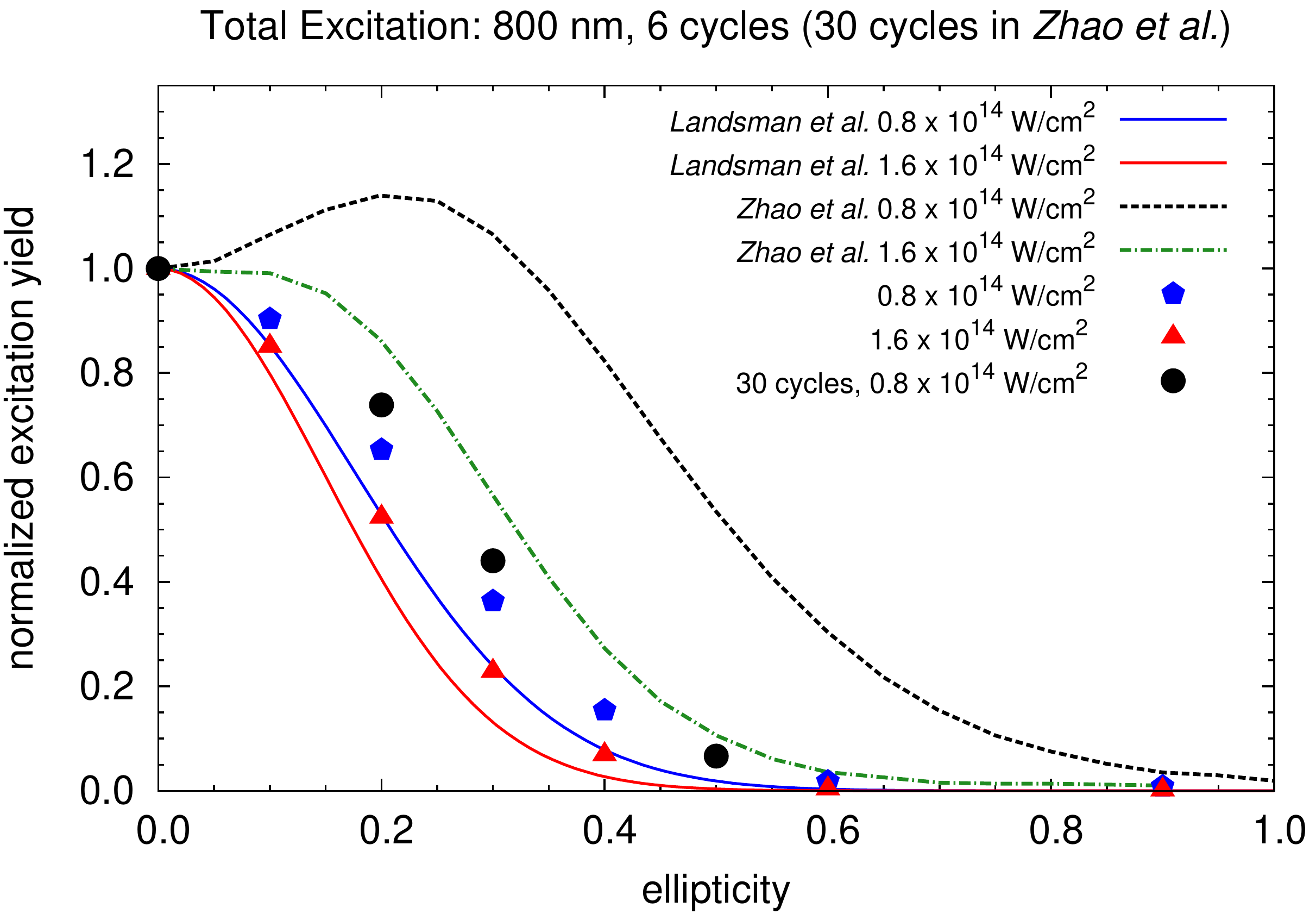}
\caption{\label{fig:fig1} Comparison of the present SAE results for total excitation with the 
predictions obtained from the Landsman {\it et al.} formula 
and those of Zhao {\it et al.} for a central wavelength of 800~nm and  
peak intensities of $\rm 0.8 \times 10^{14}\,W/cm^2$ and $\rm 1.6 \times 10^{14}\,W/cm^2$.  See text for details.}
\end{figure}

Figure~\ref{fig:fig1} shows a comparison between the present SAE results and the predictions of 
Landsman {\it et al.}~\cite{LanNJP13} and Zhao {\it et al.}~\cite{Zhao2019} for a central wavelength of 800~nm
and several peak intensities and pulse lengths.
Most of our calculations were performed for 6-cycle pulses and averaged over four CEPs, as described above.
The formula given in~\cite{LanNJP13} contains neither the pulse length nor the CEP.  The simulations in~\cite{Zhao2019}
were performed for \hbox{30-cycle}, \hbox{10-cycle}, and \hbox{4-cycle}  pulses.  
The most interesting results appears to be the one for 30 cycles, 800~nm, and a peak intensity
of  $\rm 0.8 \times 10^{14}\,W/cm^2$, where the relative FTI maximum is predicted to have an unambiguous maximum at $\varepsilon \approx 0.2$.
In order to check this, we carried out several calculations for 30 cycles, so that it was possible to obtain the relevant ratios. 
The calculations with a small but non\-zero ellipticity such
as $\varepsilon = 0.2$ are computationally demanding, and hence we do not map out the 
entire curve.  However, the available points at $\varepsilon = 0.2, 0.3,~{\rm and}~0.5$ 
clearly show that the width of the curve representing the ratio of the excited-state
population normalized to its value at $\varepsilon = 0.0$  is significantly wider for the 30~fs pulse compared to 
the 6~fs pulse at $\rm 0.8 \times 10^{14}\,W/cm^2$,
albeit not by as much as predicted by Zhao {\it et al.}~\cite{Zhao2019}.  We still find the maximum 
at $\varepsilon = 0.0$.  

The apparently better agreement of our calculations for $\rm 1.6 \times 10^{14}\,W/cm^2$ and the 
Landsman {\it et al.} formula for $\rm 0.8 \times 10^{14}\,W/cm^2$ is accidental.  It simply reflects the fact
that the Landsman {\it et al.} formula predicts the curve too narrow.
Similarly, the better agreement with Zhao {\it et al.} also for the higher intensity is due to their method
predicting the curve too wide. This clearly demonstrates  that for the processes discussed in the present paper 
neither one of the essentially classical methods is sufficiently accurate.  Instead a quantum-mechanical treatment is highly desirable. 

\begin{figure}[t!]
\includegraphics[width=0.45\textwidth]{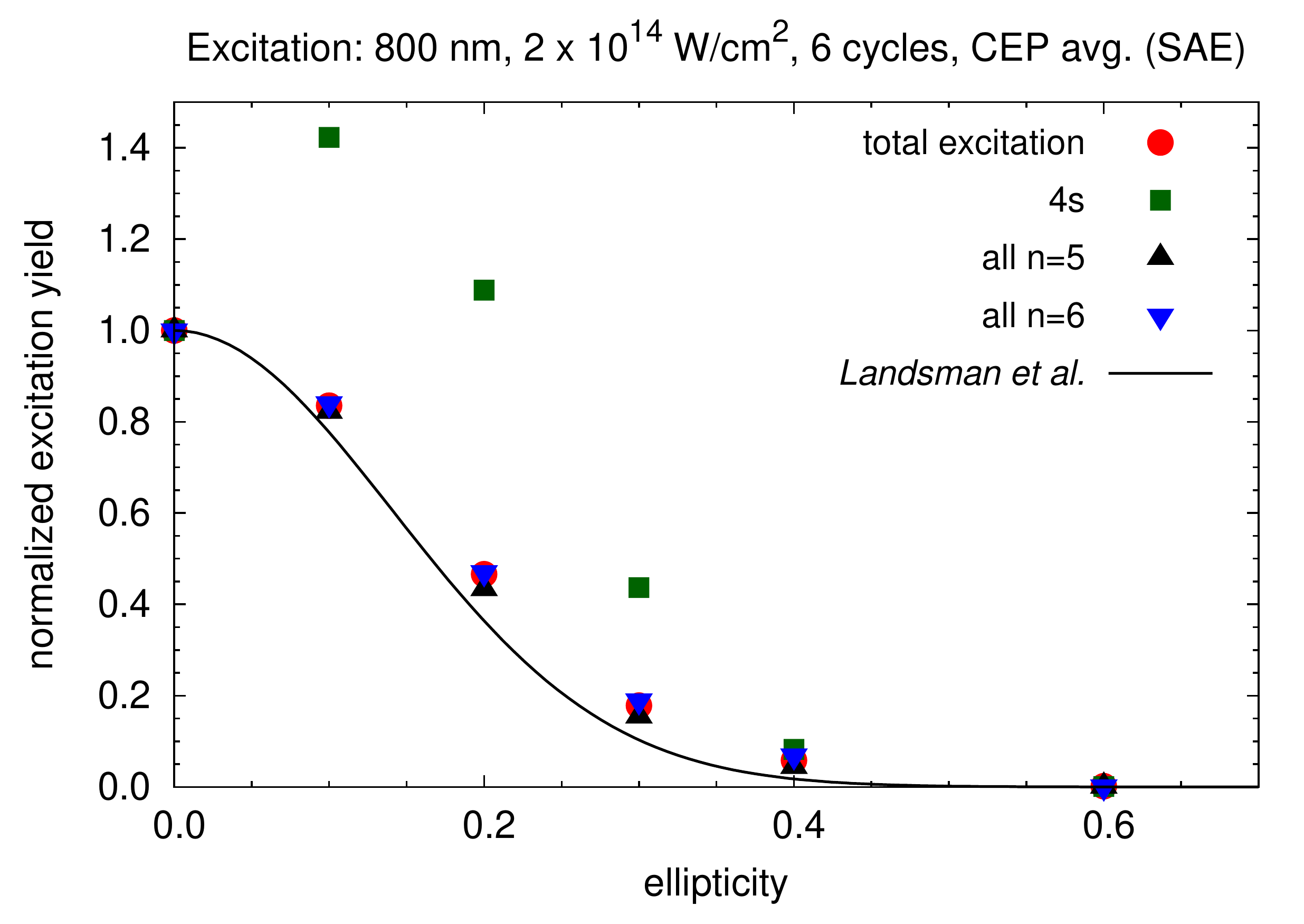}
\caption{\label{fig:fig3} CEP-averaged SAE results for total excitation of Ar($3p)$ for a 6-cycle 800~nm
pulse of peak intensity $\rm 2 \times 10^{14}\,W/cm^2$.  We present the
relative ellipticity dependence for ending up in the $4s$ orbital as well as all the $n=5$ and $n=6$ orbitals, respectively.
Also shown are the predictions from the Landsman {\it et al.} formula.
See text for details.}
\end{figure}

\begin{figure}[b!]
\includegraphics[width=0.45\textwidth]{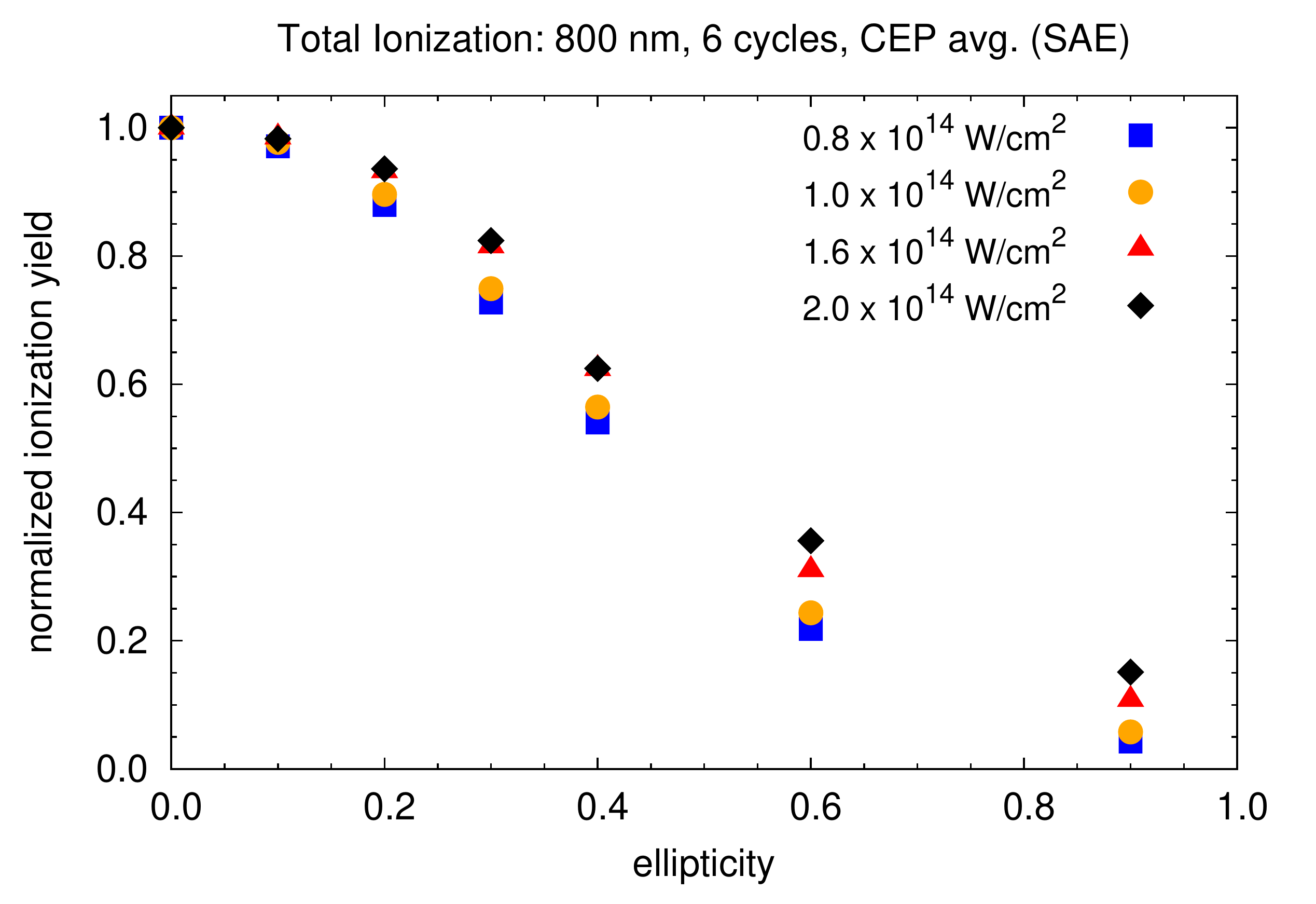}
\includegraphics[width=0.43\textwidth]{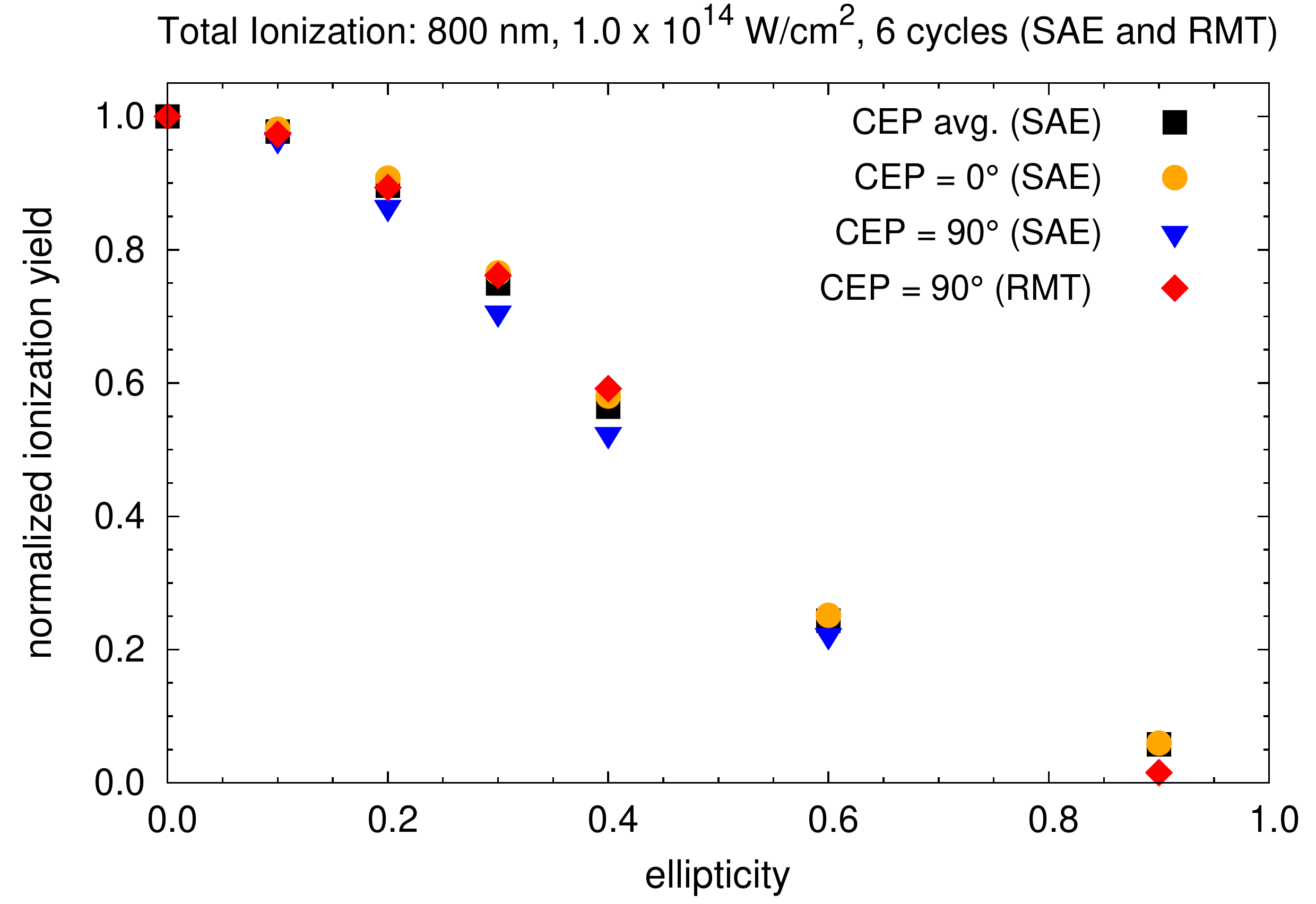}
\caption{\label{fig:fig4} Top: CEP-averaged SAE results for the relative ellipticity dependence of Ar($3p$) ionization 
for a 6-cycle 800~nm
pulse of various peak intensities indicated in the legend.  Bottom: Comparison of SAE results for ionization
for a 6-cycle 800~nm
pulse of peak intensity $\rm 10^{14}\,W/cm^2$ compared with RMT results for a few ellipticities. }
\end{figure}

Figure~\ref{fig:fig3} shows results for yet another peak intensity, this time $2 \times \rm 10^{14}\,W/cm^2$.  Here we
investigate the ellipticity dependence of ending up in the $4s$ orbital, any of the $n=5$ or $n=6$ orbitals, or any
excited  state directly at the end of the laser pulse.  Interestingly, the $4s$ excitation is predicted to peak
at a non\-zero ellipticity while $n=5$ and $n=6$ are very similar to the entire relative excitation probability.
Recall that the experimentally observed signal originates from the metastable states involving the $4s$ orbital.
However, this is not the signal that one would expect from the target immediately after the laser pulse is over but
rather much later (on an atomic timescale) when exited states have optically decayed either back to the ground state
or to either one of the $(3p^5 4s)^3P_{2,0}$ metastable states.

Finally, we show in Fig.~\ref{fig:fig4} the relative ionization probability for a number of peak intensities of
a 6-cycle 800~nm pulse obtained in the SAE model (top panel) as well as for a peak intensity of $\rm 10^{14}\,W/cm^2$ 
for two CEPs, CEP-averaged, and compared once again with RMT results for a CEP of~$90^\circ$.  Compared to the
previous figures, we clearly see a much wider curve than for FTI.  Even for $\varepsilon = 0.9$, the signal 
can visually be distinguished from zero.  Not surprisingly in light of the previous results, 
the CEP dependence is weak even for a short 6-cycle pulse.
Also, the RMT calculations once again reveal a similar ellipticity dependence as predicted by the SAE model.

\section{Conclusions}\label{sec:conclusions}
We have investigated the ellipticity dependence of 
frustrated and completed ionization of the Ar($3p^6$) ground state, when the target 
was exposed to a 6-cycle or in some cases 30-cycle 800~nm infrared pulse of with peak intensities between 
$0.8 \times \rm 10^{14}\,W/cm^2$ and $2.0 \times \rm 10^{14}\,W/cm^2$, respectively.
Predictions based on calculations in the single-active electron approximation were spot-checked against results obtained
in a multi-electron \hbox{$R$-matrix} (close-coupling) with time dependence model. Excellent agreement 
was obtained for the ratio of these probabilities as a function of ellipticity when normalized to
the results for linearly polarized light.  In practice, this ratio is close, although not identical, to what would 
be measured experimentally.  

Reasonable agreement with the analytic formula proposed by  
Landsman {\it et al.}~\cite{LanNJP13} for the relative ellipticity was obtained 
for the 6-cycle pulses. For the longer 30-cycle pulse with a peak intensity of $0.8 \times \rm 10^{14}\,W/cm^2$, the curve widens.
This is in qualitative agreement with the predictions of 
Zhao {\it et al.}~\cite{Zhao2019}, but  we do not confirm an increase at non\-zero ellipticity.  
In general, the Landsman {\it et al.} formula appears to predict the curve too narrow while
the treatment of Zhao {\it et al.} predicts it too wide.  Finally, 
we confirm that the relative ellipticity dependence for ionization follows a much wider bell curve than
that for frustrated ionization.

\bigskip

\section*{Acknowledgments}
This work was supported by the United States National Science Foundation under
grants No.~OAC-1834740 and PHY-1803844, and by the XSEDE supercomputer allocation No.~PHY-090031. 
The calculations were carried out on Stampede2 and Frontera at the Texas Advanced Computing Center,
as well as Bridges at the  Pittsburgh Supercomputer Center and Comet at the San Diego Supercomputer Center.
A.B.\ is grateful for a Michael Smith Scholarship.
X.-M.T.\ was supported by a Grant-in-Aid for Scientific Research (Grant No.\ JP16K05495) from the Japan Society for the Promotion of
Science. D.C.\ is supported by an Australian Government RTP Scholarship. 
The RMT code is part of the UK-AMOR suite.  It can be obtained free of charge from https://gitlab.com/uk-amor/rmt/.



\begin{thebibliography}{30}
\expandafter\ifx\csname natexlab\endcsname\relax\def\natexlab#1{#1}\fi
\expandafter\ifx\csname bibnamefont\endcsname\relax
  \def\bibnamefont#1{#1}\fi
\expandafter\ifx\csname bibfnamefont\endcsname\relax
  \def\bibfnamefont#1{#1}\fi
\expandafter\ifx\csname citenamefont\endcsname\relax
  \def\citenamefont#1{#1}\fi
\expandafter\ifx\csname url\endcsname\relax
  \def\url#1{\texttt{#1}}\fi
\expandafter\ifx\csname urlprefix\endcsname\relax\def\urlprefix{URL }\fi
\providecommand{\bibinfo}[2]{#2}
\providecommand{\eprint}[2][]{\url{#2}}

\bibitem[{\citenamefont{Corkum}(1993)}]{Corkum3Step}
\bibinfo{author}{\bibfnamefont{P.~B.} \bibnamefont{Corkum}},
  \bibinfo{journal}{Phys. Rev. Lett.} \textbf{\bibinfo{volume}{71}},
  \bibinfo{pages}{1994} (\bibinfo{year}{1993}).

\bibitem[{\citenamefont{Dakka et~al.}(2018)\citenamefont{Dakka, Tsiminis,
  Glover, Perrella, Moffatt, Spooner, Sang, Light, and Luiten}}]{Dakka2018}
\bibinfo{author}{\bibfnamefont{M.}~\bibnamefont{Dakka}},
  \bibinfo{author}{\bibfnamefont{G.}~\bibnamefont{Tsiminis}},
  \bibinfo{author}{\bibfnamefont{R.}~\bibnamefont{Glover}},
  \bibinfo{author}{\bibfnamefont{C.}~\bibnamefont{Perrella}},
  \bibinfo{author}{\bibfnamefont{J.}~\bibnamefont{Moffatt}},
  \bibinfo{author}{\bibfnamefont{N.}~\bibnamefont{Spooner}},
  \bibinfo{author}{\bibfnamefont{R.}~\bibnamefont{Sang}},
  \bibinfo{author}{\bibfnamefont{P.}~\bibnamefont{Light}}, \bibnamefont{and}
  \bibinfo{author}{\bibfnamefont{A.}~\bibnamefont{Luiten}},
  \bibinfo{journal}{Physical Review Letters} \textbf{\bibinfo{volume}{121}}
  (\bibinfo{year}{2018}).

\bibitem[{\citenamefont{Gao and Tong}(2019)}]{PhysRevA.100.063424}
\bibinfo{author}{\bibfnamefont{X.}~\bibnamefont{Gao}} \bibnamefont{and}
  \bibinfo{author}{\bibfnamefont{X.-M.} \bibnamefont{Tong}},
  \bibinfo{journal}{Phys. Rev. A} \textbf{\bibinfo{volume}{100}},
  \bibinfo{pages}{063424} (\bibinfo{year}{2019}).

\bibitem[{\citenamefont{Nubbemeyer et~al.}(2008)\citenamefont{Nubbemeyer,
  Gorling, Saenz, Eichmann, and Sandner}}]{PhysRevLett.101.233001}
\bibinfo{author}{\bibfnamefont{T.}~\bibnamefont{Nubbemeyer}},
  \bibinfo{author}{\bibfnamefont{K.}~\bibnamefont{Gorling}},
  \bibinfo{author}{\bibfnamefont{A.}~\bibnamefont{Saenz}},
  \bibinfo{author}{\bibfnamefont{U.}~\bibnamefont{Eichmann}}, \bibnamefont{and}
  \bibinfo{author}{\bibfnamefont{W.}~\bibnamefont{Sandner}},
  \bibinfo{journal}{Phys. Rev. Lett.} \textbf{\bibinfo{volume}{101}},
  \bibinfo{pages}{233001} (\bibinfo{year}{2008}).

\bibitem[{\citenamefont{Landsman et~al.}(2013)\citenamefont{Landsman, Pfeiffer,
  Hofmann, Smolarski, Cirelli, and Keller}}]{LanNJP13}
\bibinfo{author}{\bibfnamefont{A.~S.} \bibnamefont{Landsman}},
  \bibinfo{author}{\bibfnamefont{A.~N.} \bibnamefont{Pfeiffer}},
  \bibinfo{author}{\bibfnamefont{C.}~\bibnamefont{Hofmann}},
  \bibinfo{author}{\bibfnamefont{M.}~\bibnamefont{Smolarski}},
  \bibinfo{author}{\bibfnamefont{C.}~\bibnamefont{Cirelli}}, \bibnamefont{and}
  \bibinfo{author}{\bibfnamefont{U.}~\bibnamefont{Keller}},
  \bibinfo{journal}{New Journal of Physics} \textbf{\bibinfo{volume}{15}},
  \bibinfo{pages}{013001} (\bibinfo{year}{2013}).

\bibitem[{\citenamefont{Yun et~al.}(2018)\citenamefont{Yun, Mun, Park, Ivanov,
  Nam, and Kim}}]{Yun2018}
\bibinfo{author}{\bibfnamefont{H.}~\bibnamefont{Yun}},
  \bibinfo{author}{\bibfnamefont{J.~H.} \bibnamefont{Mun}},
  \bibinfo{author}{\bibfnamefont{S.~B.} \bibnamefont{Park}},
  \bibinfo{author}{\bibfnamefont{I.~A.} \bibnamefont{Ivanov}},
  \bibinfo{author}{\bibfnamefont{C.~H.} \bibnamefont{Nam}}, \bibnamefont{and}
  \bibinfo{author}{\bibfnamefont{K.~T.} \bibnamefont{Kim}},
  \bibinfo{journal}{Nature Photonics} \textbf{\bibinfo{volume}{12}},
  \bibinfo{pages}{620} (\bibinfo{year}{2018}).

\bibitem[{\citenamefont{Shvetsov-Shilovski
  et~al.}(2009)\citenamefont{Shvetsov-Shilovski, Goreslavski, Popruzhenko, and
  Becker}}]{Shvetsov-Shilovski2009}
\bibinfo{author}{\bibfnamefont{N.~I.} \bibnamefont{Shvetsov-Shilovski}},
  \bibinfo{author}{\bibfnamefont{S.~P.} \bibnamefont{Goreslavski}},
  \bibinfo{author}{\bibfnamefont{S.~V.} \bibnamefont{Popruzhenko}},
  \bibnamefont{and} \bibinfo{author}{\bibfnamefont{W.}~\bibnamefont{Becker}},
  \bibinfo{journal}{Laser Physics} \textbf{\bibinfo{volume}{19}},
  \bibinfo{pages}{1550 } (\bibinfo{year}{2009}).

\bibitem[{\citenamefont{Zhao et~al.}(2019)\citenamefont{Zhao, Zhou, Liang,
  Zeng, Ke, Liu, Li, and Lu}}]{Zhao2019}
\bibinfo{author}{\bibfnamefont{Y.}~\bibnamefont{Zhao}},
  \bibinfo{author}{\bibfnamefont{Y.}~\bibnamefont{Zhou}},
  \bibinfo{author}{\bibfnamefont{J.}~\bibnamefont{Liang}},
  \bibinfo{author}{\bibfnamefont{Z.}~\bibnamefont{Zeng}},
  \bibinfo{author}{\bibfnamefont{Q.}~\bibnamefont{Ke}},
  \bibinfo{author}{\bibfnamefont{Y.}~\bibnamefont{Liu}},
  \bibinfo{author}{\bibfnamefont{M.}~\bibnamefont{Li}}, \bibnamefont{and}
  \bibinfo{author}{\bibfnamefont{P.}~\bibnamefont{Lu}}, \bibinfo{journal}{Opt.\
  Express} \textbf{\bibinfo{volume}{27}}, \bibinfo{pages}{21689}
  (\bibinfo{year}{2019}).

\bibitem[{\citenamefont{Huang et~al.}(2013)\citenamefont{Huang, Xia, and
  Fu}}]{PhysRevA.87.033415}
\bibinfo{author}{\bibfnamefont{K.-y.} \bibnamefont{Huang}},
  \bibinfo{author}{\bibfnamefont{Q.-z.} \bibnamefont{Xia}}, \bibnamefont{and}
  \bibinfo{author}{\bibfnamefont{L.-B.} \bibnamefont{Fu}},
  \bibinfo{journal}{Phys. Rev. A} \textbf{\bibinfo{volume}{87}},
  \bibinfo{pages}{033415} (\bibinfo{year}{2013}).

\bibitem[{\citenamefont{Chetty et~al.}(2019)\citenamefont{Chetty, Glover, Xu,
  Palmer, deHarak, Light, Luiten, Litvinyuk, and Sang}}]{ICPEAC-Griffith}
\bibinfo{author}{\bibfnamefont{D.}~\bibnamefont{Chetty}},
  \bibinfo{author}{\bibfnamefont{R.~D.} \bibnamefont{Glover}},
  \bibinfo{author}{\bibfnamefont{H.}~\bibnamefont{Xu}},
  \bibinfo{author}{\bibfnamefont{A.~J.} \bibnamefont{Palmer}},
  \bibinfo{author}{\bibfnamefont{B.~A.} \bibnamefont{deHarak}},
  \bibinfo{author}{\bibfnamefont{P.~S.} \bibnamefont{Light}},
  \bibinfo{author}{\bibfnamefont{A.~N.} \bibnamefont{Luiten}},
  \bibinfo{author}{\bibfnamefont{I.~V.} \bibnamefont{Litvinyuk}},
  \bibnamefont{and} \bibinfo{author}{\bibfnamefont{R.~T.} \bibnamefont{Sang}},
  in \emph{\bibinfo{booktitle}{Book of Abstracts: XXXI International Conference
  on Photonic, Electronic, and Atomic Collisions}}
  (\bibinfo{organization}{ICPEAC}, \bibinfo{year}{2019}),
  p.~\bibinfo{pages}{81}.

\bibitem[{\citenamefont{Chetty et~al.}(2020)\citenamefont{Chetty, Glover,
  deHarak, Tong, Xu, Pauly, Smith, Hamilton, Bartschat, Ziegel
  et~al.}}]{fti_linear_ar}
\bibinfo{author}{\bibfnamefont{D.}~\bibnamefont{Chetty}},
  \bibinfo{author}{\bibfnamefont{R.~D.} \bibnamefont{Glover}},
  \bibinfo{author}{\bibfnamefont{B.~A.} \bibnamefont{deHarak}},
  \bibinfo{author}{\bibfnamefont{X.~M.} \bibnamefont{Tong}},
  \bibinfo{author}{\bibfnamefont{H.}~\bibnamefont{Xu}},
  \bibinfo{author}{\bibfnamefont{T.}~\bibnamefont{Pauly}},
  \bibinfo{author}{\bibfnamefont{N.}~\bibnamefont{Smith}},
  \bibinfo{author}{\bibfnamefont{K.~R.} \bibnamefont{Hamilton}},
  \bibinfo{author}{\bibfnamefont{K.}~\bibnamefont{Bartschat}},
  \bibinfo{author}{\bibfnamefont{J.~P.} \bibnamefont{Ziegel}},
  \bibnamefont{et~al.}, \bibinfo{journal}{Phys. Rev. A}
  \textbf{\bibinfo{volume}{101}}, \bibinfo{pages}{053402}
  (\bibinfo{year}{2020}).

\bibitem[{\citenamefont{Tong and Lin}(2005)}]{Tong05c}
\bibinfo{author}{\bibfnamefont{X.~M.} \bibnamefont{Tong}} \bibnamefont{and}
  \bibinfo{author}{\bibfnamefont{C.~D.} \bibnamefont{Lin}},
  \bibinfo{journal}{J. Phys. B: At. Mol. Opt. Phys.}
  \textbf{\bibinfo{volume}{38}}, \bibinfo{pages}{2593} (\bibinfo{year}{2005}).

\bibitem[{\citenamefont{Wragg et~al.}(2019)\citenamefont{Wragg, Clarke,
  Armstrong, Brown, Ballance, and van~der Hart}}]{spin-orbit-rmt}
\bibinfo{author}{\bibfnamefont{J.}~\bibnamefont{Wragg}},
  \bibinfo{author}{\bibfnamefont{D.~D.~A.} \bibnamefont{Clarke}},
  \bibinfo{author}{\bibfnamefont{G.~S.~J.} \bibnamefont{Armstrong}},
  \bibinfo{author}{\bibfnamefont{A.~C.} \bibnamefont{Brown}},
  \bibinfo{author}{\bibfnamefont{C.~P.} \bibnamefont{Ballance}},
  \bibnamefont{and} \bibinfo{author}{\bibfnamefont{H.~W.} \bibnamefont{van~der
  Hart}}, \bibinfo{journal}{Phys. Rev. Lett.} \textbf{\bibinfo{volume}{123}},
  \bibinfo{pages}{163001} (\bibinfo{year}{2019}).

\bibitem[{\citenamefont{Ivanov et~al.}(2014)\citenamefont{Ivanov, Kheifets,
  Bartschat, Emmons, Buczek, Gryzlova, and
  Grum-Grzhimailo}}]{PhysRevA.90.043401}
\bibinfo{author}{\bibfnamefont{I.~A.} \bibnamefont{Ivanov}},
  \bibinfo{author}{\bibfnamefont{A.~S.} \bibnamefont{Kheifets}},
  \bibinfo{author}{\bibfnamefont{K.}~\bibnamefont{Bartschat}},
  \bibinfo{author}{\bibfnamefont{J.}~\bibnamefont{Emmons}},
  \bibinfo{author}{\bibfnamefont{S.~M.} \bibnamefont{Buczek}},
  \bibinfo{author}{\bibfnamefont{E.~V.} \bibnamefont{Gryzlova}},
  \bibnamefont{and} \bibinfo{author}{\bibfnamefont{A.~N.}
  \bibnamefont{Grum-Grzhimailo}}, \bibinfo{journal}{Phys. Rev. A}
  \textbf{\bibinfo{volume}{90}}, \bibinfo{pages}{043401}
  (\bibinfo{year}{2014}).

\bibitem[{\citenamefont{Crank and Nicolson}(1996)}]{CN1996}
\bibinfo{author}{\bibfnamefont{J.}~\bibnamefont{Crank}} \bibnamefont{and}
  \bibinfo{author}{\bibfnamefont{P.}~\bibnamefont{Nicolson}},
  \bibinfo{journal}{Advances in Computational Mathematics}
  \textbf{\bibinfo{volume}{6}}, \bibinfo{pages}{207} (\bibinfo{year}{1996}).

\bibitem[{\citenamefont{Nurhuda and Faisal}(1999)}]{NF1999}
\bibinfo{author}{\bibfnamefont{M.}~\bibnamefont{Nurhuda}} \bibnamefont{and}
  \bibinfo{author}{\bibfnamefont{F.~H.~M.} \bibnamefont{Faisal}},
  \bibinfo{journal}{Phys. Rev. A} \textbf{\bibinfo{volume}{60}},
  \bibinfo{pages}{3125} (\bibinfo{year}{1999}).

\bibitem[{\citenamefont{Douguet et~al.}(2016)\citenamefont{Douguet,
  Grum-Grzhimailo, Gryzlova, Staroselskaya, Venzke, and
  Bartschat}}]{PhysRevA.93.033402}
\bibinfo{author}{\bibfnamefont{N.}~\bibnamefont{Douguet}},
  \bibinfo{author}{\bibfnamefont{A.~N.} \bibnamefont{Grum-Grzhimailo}},
  \bibinfo{author}{\bibfnamefont{E.~V.} \bibnamefont{Gryzlova}},
  \bibinfo{author}{\bibfnamefont{E.~I.} \bibnamefont{Staroselskaya}},
  \bibinfo{author}{\bibfnamefont{J.}~\bibnamefont{Venzke}}, \bibnamefont{and}
  \bibinfo{author}{\bibfnamefont{K.}~\bibnamefont{Bartschat}},
  \bibinfo{journal}{Phys. Rev. A} \textbf{\bibinfo{volume}{93}},
  \bibinfo{pages}{033402} (\bibinfo{year}{2016}).

\bibitem[{\citenamefont{Brown et~al.}(2020)\citenamefont{Brown, Armstrong,
  Benda, Clarke, Wragg, Hamilton, Mašín, Gorfinkiel, and van~der
  Hart}}]{rmt_cpc}
\bibinfo{author}{\bibfnamefont{A.~C.} \bibnamefont{Brown}},
  \bibinfo{author}{\bibfnamefont{G.~S.} \bibnamefont{Armstrong}},
  \bibinfo{author}{\bibfnamefont{J.}~\bibnamefont{Benda}},
  \bibinfo{author}{\bibfnamefont{D.~D.} \bibnamefont{Clarke}},
  \bibinfo{author}{\bibfnamefont{J.}~\bibnamefont{Wragg}},
  \bibinfo{author}{\bibfnamefont{K.~R.} \bibnamefont{Hamilton}},
  \bibinfo{author}{\bibfnamefont{Z.}~\bibnamefont{Mašín}},
  \bibinfo{author}{\bibfnamefont{J.~D.} \bibnamefont{Gorfinkiel}},
  \bibnamefont{and} \bibinfo{author}{\bibfnamefont{H.~W.} \bibnamefont{van~der
  Hart}}, \bibinfo{journal}{Computer Physics Communications}
  \textbf{\bibinfo{volume}{250}}, \bibinfo{pages}{107062}
  (\bibinfo{year}{2020}), ISSN \bibinfo{issn}{0010-4655}.

\bibitem[{\citenamefont{Clarke et~al.}(2018{\natexlab{a}})\citenamefont{Clarke,
  Armstrong, Brown, and van~der Hart}}]{rmt_clarke}
\bibinfo{author}{\bibfnamefont{D.~D.~A.} \bibnamefont{Clarke}},
  \bibinfo{author}{\bibfnamefont{G.~S.~J.} \bibnamefont{Armstrong}},
  \bibinfo{author}{\bibfnamefont{A.~C.} \bibnamefont{Brown}}, \bibnamefont{and}
  \bibinfo{author}{\bibfnamefont{H.~W.} \bibnamefont{van~der Hart}},
  \bibinfo{journal}{Phys. Rev. A} \textbf{\bibinfo{volume}{98}},
  \bibinfo{pages}{053442} (\bibinfo{year}{2018}{\natexlab{a}}).

\bibitem[{\citenamefont{Moore et~al.}(2011)\citenamefont{Moore, Lysaght,
  Nikolopoulos, Parker, van~der Hart, and Taylor}}]{rmt_moore}
\bibinfo{author}{\bibfnamefont{L.~R.} \bibnamefont{Moore}},
  \bibinfo{author}{\bibfnamefont{M.~A.} \bibnamefont{Lysaght}},
  \bibinfo{author}{\bibfnamefont{L.~A.~A.} \bibnamefont{Nikolopoulos}},
  \bibinfo{author}{\bibfnamefont{J.~S.} \bibnamefont{Parker}},
  \bibinfo{author}{\bibfnamefont{H.~W.} \bibnamefont{van~der Hart}},
  \bibnamefont{and} \bibinfo{author}{\bibfnamefont{K.~T.}
  \bibnamefont{Taylor}}, \bibinfo{journal}{J. Mod. Optics}
  \textbf{\bibinfo{volume}{58}}, \bibinfo{pages}{1132} (\bibinfo{year}{2011}).

\bibitem[{\citenamefont{Burke}(2011)}]{burke}
\bibinfo{author}{\bibfnamefont{P.~G.} \bibnamefont{Burke}},
  \emph{\bibinfo{title}{R-matrix Theory of Atomic Collisions}}
  (\bibinfo{publisher}{Springer Berlin Heidelberg}, \bibinfo{year}{2011}).

\bibitem[{\citenamefont{Burke and Taylor}(1975)}]{rmt_argon_structure}
\bibinfo{author}{\bibfnamefont{P.~G.} \bibnamefont{Burke}} \bibnamefont{and}
  \bibinfo{author}{\bibfnamefont{K.~T.} \bibnamefont{Taylor}},
  \bibinfo{journal}{J. Phys. B. At. Mol. Opt. Phys.}
  \textbf{\bibinfo{volume}{8}}, \bibinfo{pages}{2620} (\bibinfo{year}{1975}).

\bibitem[{\citenamefont{Hutchinson et~al.}(2010)\citenamefont{Hutchinson,
  Lysaght, and van~der Hart}}]{tdrm_dipole_gauge}
\bibinfo{author}{\bibfnamefont{S.}~\bibnamefont{Hutchinson}},
  \bibinfo{author}{\bibfnamefont{M.~A.} \bibnamefont{Lysaght}},
  \bibnamefont{and} \bibinfo{author}{\bibfnamefont{H.~W.} \bibnamefont{van~der
  Hart}}, \bibinfo{journal}{J. Phys. B. At. Mol. Opt. Phys.}
  \textbf{\bibinfo{volume}{43}}, \bibinfo{pages}{095603}
  (\bibinfo{year}{2010}).

\bibitem[{\citenamefont{Arnoldi}(1951)}]{arnoldi}
\bibinfo{author}{\bibfnamefont{W.~E.} \bibnamefont{Arnoldi}},
  \bibinfo{journal}{The Quarterly Journal of Pure and Applied Mathematics}
  \textbf{\bibinfo{volume}{9}}, \bibinfo{pages}{17} (\bibinfo{year}{1951}).

\bibitem[{\citenamefont{Hamilton et~al.}(2017)\citenamefont{Hamilton, van~der
  Hart, and Brown}}]{hamilton_two_colour}
\bibinfo{author}{\bibfnamefont{K.~R.} \bibnamefont{Hamilton}},
  \bibinfo{author}{\bibfnamefont{H.~W.} \bibnamefont{van~der Hart}},
  \bibnamefont{and} \bibinfo{author}{\bibfnamefont{A.~C.} \bibnamefont{Brown}},
  \bibinfo{journal}{Phys. Rev. A} \textbf{\bibinfo{volume}{95}},
  \bibinfo{pages}{013408} (\bibinfo{year}{2017}).

\bibitem[{\citenamefont{Hassouneh et~al.}(2014)\citenamefont{Hassouneh, Brown,
  and van~der Hart}}]{ola_nearIR}
\bibinfo{author}{\bibfnamefont{O.}~\bibnamefont{Hassouneh}},
  \bibinfo{author}{\bibfnamefont{A.~C.} \bibnamefont{Brown}}, \bibnamefont{and}
  \bibinfo{author}{\bibfnamefont{H.~W.} \bibnamefont{van~der Hart}},
  \bibinfo{journal}{Phys. Rev. A} \textbf{\bibinfo{volume}{90}},
  \bibinfo{pages}{043418} (\bibinfo{year}{2014}).

\bibitem[{\citenamefont{Brown and van~der Hart}(2016)}]{brown_xuvhhg}
\bibinfo{author}{\bibfnamefont{A.~C.} \bibnamefont{Brown}} \bibnamefont{and}
  \bibinfo{author}{\bibfnamefont{H.~W.} \bibnamefont{van~der Hart}},
  \bibinfo{journal}{Phys. Rev. Lett.} \textbf{\bibinfo{volume}{117}},
  \bibinfo{pages}{093201} (\bibinfo{year}{2016}).

\bibitem[{\citenamefont{Clarke et~al.}(2018{\natexlab{b}})\citenamefont{Clarke,
  van~der Hart, and Brown}}]{clarke_xihhg}
\bibinfo{author}{\bibfnamefont{D.~D.~A.} \bibnamefont{Clarke}},
  \bibinfo{author}{\bibfnamefont{H.~W.} \bibnamefont{van~der Hart}},
  \bibnamefont{and} \bibinfo{author}{\bibfnamefont{A.~C.} \bibnamefont{Brown}},
  \bibinfo{journal}{Phys. Rev. A} \textbf{\bibinfo{volume}{97}},
  \bibinfo{pages}{023413} (\bibinfo{year}{2018}{\natexlab{b}}).

\bibitem[{\citenamefont{Hassouneh et~al.}(2015)\citenamefont{Hassouneh, Law,
  Shearer, Brown, and van~der Hart}}]{ola_rescattering}
\bibinfo{author}{\bibfnamefont{O.}~\bibnamefont{Hassouneh}},
  \bibinfo{author}{\bibfnamefont{S.}~\bibnamefont{Law}},
  \bibinfo{author}{\bibfnamefont{S.~F.~C.} \bibnamefont{Shearer}},
  \bibinfo{author}{\bibfnamefont{A.~C.} \bibnamefont{Brown}}, \bibnamefont{and}
  \bibinfo{author}{\bibfnamefont{H.~W.} \bibnamefont{van~der Hart}},
  \bibinfo{journal}{Phys. Rev. A} \textbf{\bibinfo{volume}{91}},
  \bibinfo{pages}{031404} (\bibinfo{year}{2015}).

\bibitem[{\citenamefont{Levy and Huffman}(1969)}]{LEVY19691349}
\bibinfo{author}{\bibfnamefont{M.}~\bibnamefont{Levy}} \bibnamefont{and}
  \bibinfo{author}{\bibfnamefont{R.}~\bibnamefont{Huffman}},
  \bibinfo{journal}{Journal of Quantitative Spectroscopy and Radiative
  Transfer} \textbf{\bibinfo{volume}{9}}, \bibinfo{pages}{1349 }
  (\bibinfo{year}{1969}).

\end{thebibliography}
\end{document}